\documentclass[useAMS,usenatbib]{mn2e}

\usepackage{amsmath}
\usepackage[final]{graphicx}
\usepackage{wasysym}
\usepackage{amssymb}
\usepackage{subfigure}
\usepackage[T1]{fontenc}
\usepackage{aecompl}
\newcommand*{\myalign}[2]{\multicolumn{1}{#1}{#2}}

\title[Solar analogues from HARPS]{Solar analogues and Solar twins in the HARPS
archive
  \thanks{Based on observations from the ESO Science Archive Facility under request datson34792 and following.}}

\author[Datson et al.]{Juliet Datson$^{1}$\thanks{E-mail: juliet.datson@utu.fi},
  Chris Flynn$^{2,3,4}$ and Laura Portinari$^{1}$\\ 
$^{1}$Tuorla Observatory, Department of Physics and Astronomy,
University of Turku, Finland,\\
$^{2}$Centre for Astrophysics and Supercomputing, Swinburne University of Technology, VIC 3122 Australia\\
$^{3}$Finnish Centre for Astronomy with ESO, University of Turku, FI-21500, Piikkio, Finland\\
$^{4}$Department of Physics and Astronomy, University of Sydney, NSW 2006 Australia}

\begin{document}

\date{Accepted 2014 January 6.  Received 2013 December 21; in original form 2013 November 5th}

\pagerange{\pageref{firstpage}--\pageref{lastpage}} \pubyear{2013}

\maketitle

\label{firstpage}

\begin{abstract}
We present 63 Solar analogues and twins for which high S/N archival data
are available for the HARPS high resolution spectrograph at the ESO 3.6m
telescope. We perform a differential analysis of these stellar spectra relative
to the Solar spectrum, similar to previous work using ESO 2.2m/FEROS data, and
expand our analysis by introducing a new method to test the temperature
and metallicity calibration of Sun-like stars in the Geneva-Copenhagen-Survey
(GCS). The HARPS data are significantly better than the FEROS data, with
improvements in S/N, spectral resolution, and number of lines we can
analyse. We confirm the offsets to the photometric scale found in our FEROS
study.  We confirm 3 Solar twins found in the FEROS data as Solar twins in the
HARPS data, as well as identify 6 new twins.

\end{abstract}


\section{Introduction}
\label{sec:intro}

This is the era of large stellar surveys of the Milky Way, with the aim of
producing homogeneous catalogues of the kinematical and physical properties of
very large numbers of stars, from $\sim 10^5$ in HERMES/GALAH \citep{b52} and
the Gaia-ESO \citep{b53} survey, through to the $\sim 10^9$ stars to be
observed by GAIA \citep[GAIA, ][]{b44}. The basis for this ambitious surveying
has its roots in the highly successful HIPPARCOS mission in the 1990s
\citep{b47,b26}, and 2MASS \citep[2MASS, ][]{b42}, and SDSS \citep[SDSS,
][]{b41} in the early 2000s.

The estimation of precise and accurate physical parameters for these stars is
crucially important for the investigation of a wide range of questions about
the Milky Way, such as its chemical evolution, the stellar mass and luminosity
functions, secular heating and radial migration of stars in the disk, and its
star formation history.
 
The Geneva--Copenhagen Survey \citep*[GCS, ][]{b9,b27,b21} is the largest effort
to date in this direction. Surveying some 14,000 F, G and K dwarf stars in the
Solar neighbourhood, it has been used to examine a wide range of problems --
from determining structures in the Milky Way \citep*[e.g.][]{b57,b55}, over
abundance analysis of different stellar populations \citep[e.g.][]{b54}, all
the way to characterising planet host stars \citep[e.g.][]{b56}.  The stellar
physical parameters, primarily effective temperature, metallicity and stellar
ages are estimated from parallaxes and a careful calibration of the
Str{\"o}mgren indices. Since different photometric (or
spectroscopic) techniques can differ
by up to 100~K in temperature and 0.1~dex in metallicity,
it is important to have independent ways to
gauge a calibration.

In previous work, \citet*{b38} (Paper~I), inspired by standard techniques
to select spectroscopic solar twins, we introduced new methods to test a
temperature and metallicity scale with respect to the solar pinpoint.
We performed a differential comparison
of the strength of 109 weak, isolated lines in $\sim$100 stars, selected to be
as Sun-like as possible from the GCS catalog, relative to a Solar reflection
spectrum from the asteroid Ceres. We found, using FEROS data taken at the
ESO/MPIA 2.2m telescope, that the GCS temperature and metallicity scales appear
offset relative to the Sun by $-97 \pm 35$~K and $-0.12 \pm 0.02$~dex,
respectively, similar to a finding made independently by \citet{b24}.

Very high S/N and high resolution data for many of the GCS stars are available
in the HARPS \citep{b39} archive of the ESO 3.6m telescope, taken as part of
the programme to measure radial velocities of bright F, G and K stars in the
search for stars which host planets. 

In this paper, we have used HARPS data for 63 GCS stars found in the archive,
to examine with high precision this potential offset in the GCS temperature and
metallicity scale. We introduce a new method to measure the offsets,
as well as adapt some of the same methods we developed for Paper~I to this
higher resolution and much higher S/N data. We anticipate that, although we
have so far focussed on the GCS catalogue, the differential comparison approach
we have developed can be used to test the zero--point of the temperature and
metallicity scale in any survey.

This large sample of Solar analogues in the HARPS archive also allowed us to
further search for Solar twins --- the best ones identified to date being 18
Sco (HD146233) and HIP56948 (HD101364) \citep[][Paper I]{b33,b45,b11,b46} ---
which we determined through the use of some of the methods we developed in
Paper~I, but also introduce a new method; all are based on differential comparisons
of our target stars to an asteroid spectrum, which serves as our stand-in Solar
spectrum.

We organise the paper as follows: in section~\ref{sec:data} we describe how we
selected Solar analogues in the HARPS archive and describe the data reduction.
In section~\ref{sec:analysis} we present our analysis procedure to test the GCS
catalogue calibration: a new method to compare stellar versus Solar spectra,
along with our previous methods from Paper~I. In section~\ref{sec:twins} we
then use these methods to find Solar twins in our data.  Finally in
section~\ref{sec:sum} we summarise and draw our conclusions.

\section[]{Candidate selection and data reduction}
\label{sec:data}

We have selected our Solar analogues, as we did in Paper~I, by choosing stars from
the GCS-III \citep{b21} (i.e. the latest version of the catalogue) which
bracket the Solar colour $(b-y)$ in the Str\"omgren system,
absolute visual magnitude $M_{V}$ and metallicity. We adopted
the Solar values of $(b-y)_\odot = 0.403$ \citep*{b19},
$M_{V} = 4.83$ \citep{b22} and [Fe/H] $=0.0$ (by definition),
and select stars in the ranges: $0.371 < (b-y) < 0.435$ in
colour, $4.63 < M_{V} < 5.03$ in absolute magnitude and $-0.25
< [\mathrm{Fe/H}] < 0.15$ in metallicity. Note that the metallicity window is
asymmetric, extended at the metal--poor end, to allow for the possible
metallicity offset of $-0.1$~dex in the GCS scale, as found in Paper~I.

We have searched the data archive of the High Accuracy Radial velocity Planet
Searcher (HARPS) \citep{b39} at the ESO 3.6m telescope at La Silla, finding 85
of our GCS stars which satisfy the above criteria and for which there is publicly
accessible data in HARPS. We excluded two stars with highly broadened spectral
lines (HD45270 and HD118072) which made it difficult for us to compare them to
our Solar spectrum. These stars have very high rotation values, their
$v$sin$i$ being 17.6 km/s and 21.4 km/s respectively \citep{b49}, whereas the
Sun's value is only 2.29 km/s \citep{b48}. This left us with a sample of 83 stars, most of
which had been observed many times with HARPS, allowing us to estimate our
internal errors very well. We obtained a total of 408 spectra for our 83 stars,
with signal--to--noise (S/N) values in the range 30 to 450.

The spectra were acquired from 2003 to 2011. HARPS is an echelle spectrograph,
with a resolution of R$\sim$115,000 and covering a spectral range of
3780--6910\AA. As a consequence, HARPS covers a smaller wavelength range than
our FEROS (3500--9200\AA) observations in Paper~I, but offers more than twice
the resolution of FEROS (which is R$\sim$48,000). This allowed us to triple the
number of atomic lines compared to what we used in Paper I.

We found a single Ceres spectrum for use as our reflected Solar spectrum in the
HARPS archive. We found other asteroid spectra as well as the Jovian moon
Ganymede in the archive. Comparing these spectra with Ceres showed that there
were no systematic differences in the equivalent widths of the spectral lines:
in other words, they yield good Solar reflection spectra too, for all spectral
lines used in our analysis. The Ceres spectrum was the best in terms of having
the highest S/N ratio ($\sim$220) and is our Solar standard throughout this
paper.

An asteroid or Jovian moon necessarily reflects the spectrum of the Sun
close to its equatorial plane, whereas stars are observed at
random angles to this plane. Reassuringly, \citet{b59} have shown that
this has no significant effect on the equivalent widths of the observed
spectral lines, by comparing those determined from the Sun, taken at
different position angles.

The HARPS pipeline provides what are essentially science ready spectra. They
had been corrected for the effects of the CCD bias, dark current, flatfielding
and cosmic rays. The spectra were then extracted and rebinned to
0.01\AA\ linear resolution. The wavelength calibration is based on Thorium
lines from arcs, taken each afternoon.

The extracted 1-D spectra showed broad wiggles in the continuum levels. These
could be quite straightforwardly corrected out using the same approach as in
Paper~I for similar wiggles in the FEROS data. We made piecewise
estimates of the continuum levels in 10-\AA\ sections of the spectra and
normalised them by linear fitting. This resulted, as in Paper I, in very flat
spectra, in particular around the weak spectral lines for which we aimed to
measure equivalent widths.


\section{Differential Analysis of Equivalent Widths}
\label{sec:analysis}

Our analysis is based on a differential comparison between the spectra of the
target stars and the reference Solar spectrum. We used the same program,
TWOSPEC, as in Paper~I, to measure equivalent widths of lines for a wide range
of species, both neutral and ionised, and measure the median difference for
these species in the star compared to the Sun (see Paper~I for full details on
TWOSPEC). We adapted the line list from \citet*{b51} for our resolution and
wavelength range for use with our HARPS data. This resulted in a list three
times as long as the one used in Paper~I,  i.e. 321 weak, unblended lines
from ten different elements (Na, Mg, Al, Si, Ca, Ti, Cr, Fe, Ni and Zn), six of which
have lines from the ionised as well as the neutral state.
The list is available from the authors upon request.

In most of what follows we use the same methods as described in Paper~I (but
also introduce a new method -- see section \ref{sub:nim}) to test the temperature 
and metallicity calibration in the GCS catalogue for
Solar-type stars (this section) and to search for new twins (section \ref{sec:twins}).

\begin{figure}
  \includegraphics[width=90mm]{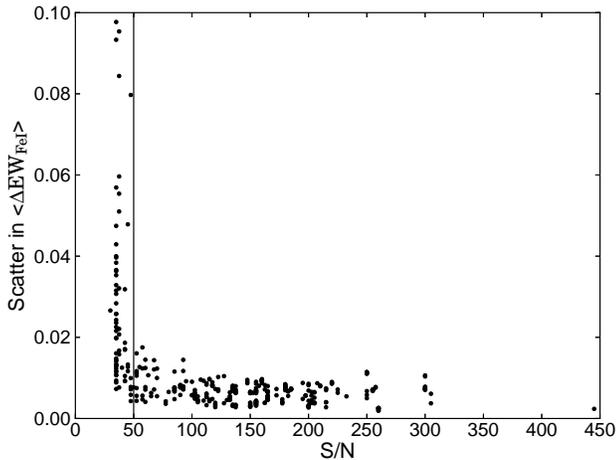}
  \caption[Optional caption for list of figures]{The scatter in the median
    difference in EW relative to the Sun, for all stars in our sample, as a
    function of the spectrum S/N. The scatter increases very significantly for
    S/N below 50 (solid line). These spectra have been dropped from our
    analysis.}
  \label{fig:SN}
\end{figure}

As before, our approach is to measure the median difference in equivalent width
(EW) of selected lines between the star's and the Solar reference spectrum,
relative to the EW in the Sun. For a Solar twin, this quantity will vanish (to
within the errors).
\begin{equation} \label{eq:meandiff}
  \textless\Delta \mathrm{EW} \textgreater=
  \textless \frac{(\mathrm{EW}(\star)-\mathrm{EW}
  (\astrosun))}{\mathrm{EW}(\astrosun)} \textgreater
\end{equation}
We begin by examining the S/N distribution of our spectra and its effect on our
measure of this quantity. This is shown in Fig.~\ref{fig:SN}. For the test we
considered Fe\thinspace I lines, which are the most numerous in our line list.

Below a S/N of approximately 50, the scatter in the medians for multiple
observations of the same star increases very significantly, and all spectra
with S/N$<50$ were removed from the sample. We have checked and confirm that
the results of the paper do not change significantly if we impose a stricter
S/N selection.

This culling left us with 329 spectra with $50< $ S/N $ <450$, for a total of 63
individual stars, of which 15 have been also observed in our FEROS sample of
Paper I (including 3 Solar twins) and thus allow a comparison with our earlier
results.

\subsection{The Neutral-Ionised (n/i) Method}
\label{sub:nim}

One way to disentangle temperature, metallicity and gravity effects on the lines
in the spectra of Sun--like stars is to exploit the different sensitivities of
the neutral versus the ionised lines of the same element.

We have searched the literature for lines of both ionisation states (neutral
and singly ionised) for as many species as possible for inclusion in our line
list. The main source of lines is the work of \citet{b51}, and includes reasonable 
numbers of lines from Fe, Ti, Ca, Ni, Cr
and Si. We carefully examined our Ceres and other HARPS spectra, as we had
done for FEROS, to remove any lines which were too weak (generally those with
EWs $< 15$ m\AA) or had nearby lines which would make EW measurements
unreliable. Far more lines could be used than for our FEROS data, as the
spectral resolution of HARPS is a factor of more than 2 greater, and the final
list contained 323 lines. Table~\ref{tab:lines} shows the final numbers of
lines we have used for elements with lines in both ionisation states.

\begin{table}
 \centering
     \caption{Numbers of lines used in for each species in our
       neutral-ionised-method for finding Sun-like stars. The lines lie in the
       HARPS spectral range.}
     \label{tab:lines}
  \begin{tabular}{@{}lcc@{}}
\hline
Element & number of neutral lines & number of ionised lines \\
\hline
Iron     & 126 & 25 \\
Titanium &  19 & 17 \\
Calcium  &  22 &  4 \\
Nickel   &  42 &  5 \\
Chromium &   9 &  3 \\
Silicon  &  15 & 20 \\
\hline
\end{tabular}
\end{table}

\subsubsection{The technique}
\label{subsub:method}

We ran TWOSPEC to compare the EWs of each element, in the stellar vs.\ Ceres
spectrum, separating the neutral and ionized lines, and define a median
difference in EW for every element and each state:

\begin{equation} \label{eq:meandiff2}
  \textless\Delta \mathrm{EW}_{\mathrm{n/i}} \textgreater=
  \textless \frac{(\mathrm{EW}_{\mathrm{n/i}}(\star)-\mathrm{EW}_{\mathrm{n/i}}
  (\astrosun))}{\mathrm{EW}_{\mathrm{n/i}}(\astrosun)} \textgreater
\end{equation}

with the subscripts n indicating neutral elements and i ionised elements.

\begin{figure}
  \centering
  \includegraphics[scale=0.4]{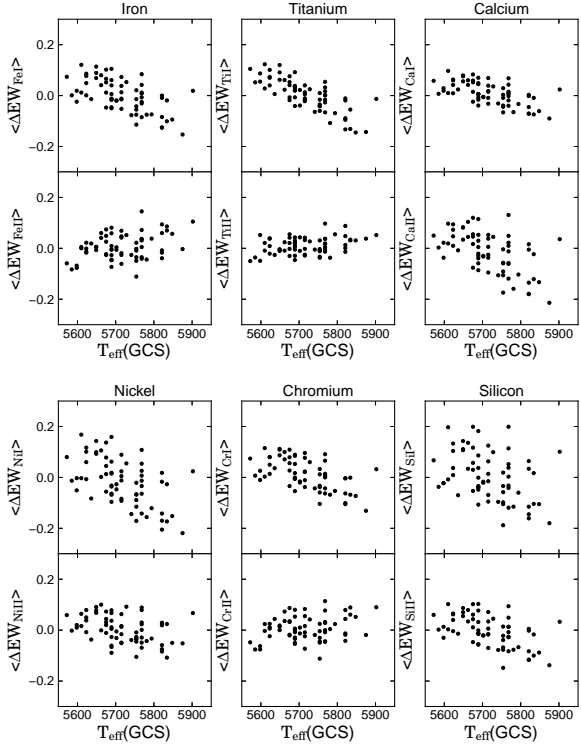}
  \caption[Optional caption for list of figures]{The median relative difference
    in equivalent widths of the spectral lines as a function of the stellar
    temperature from the GCS catalogue, for Fe, Ti, Ca, Ni, Cr and Si. Upper
    panels in each plot show result for the neutral lines, and lower panels for
    the singly ionised lines. Correlations, anti-correlations and weak
    correlations are all seen in the plot. We exploit these different trends to
    probe for any offset in the temperature and metalicity scales of GCS
    relative to the Sun, and to isolate the most-like stars in the sample.}
  \label{fig:temp}
\end{figure}

\begin{figure}
  \centering 
  \includegraphics[scale=0.4]{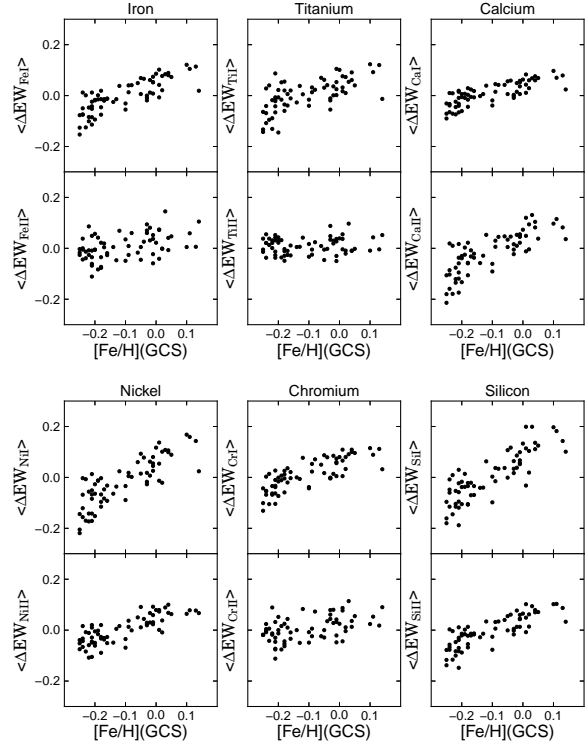}
  \caption[Optional caption for list of figures]{As for figure \ref{fig:temp}, but showing
    the trends versus the GCS metallicity of the stars.}
  \label{fig:metal}
\end{figure}

\begin{figure}
  \centering \rotatebox{+1}{\includegraphics[scale=0.4, angle=-1]{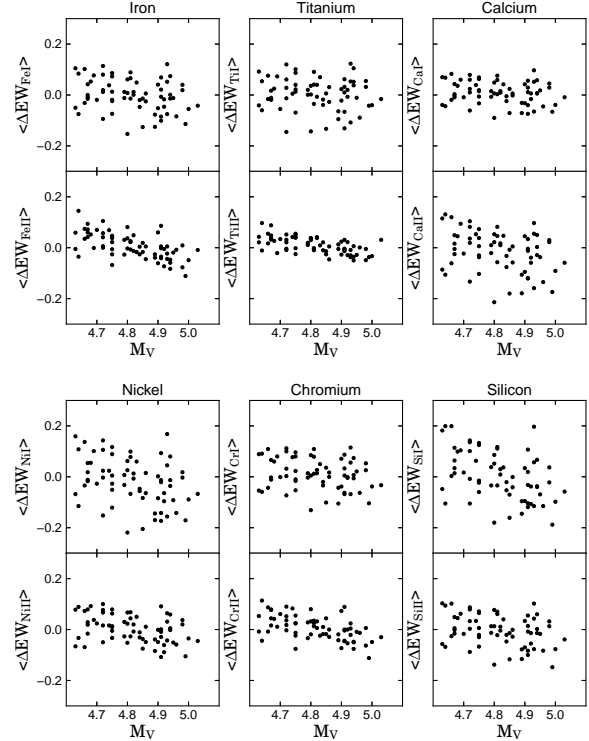}}
  \caption[Optional caption for list of figures]{As for figure \ref{fig:temp}, but showing
    trends versus the absolute magnitude, $M_{V}$ of the stars.}
  \label{fig:mv}
\end{figure}

In Figs.~\ref{fig:temp},~\ref{fig:metal} and \ref{fig:mv} , we show
the median EW differences relative to the Sun for all stars in the sample,
plotted as a function of the GCS values of $T_{\mathrm{eff}}$, [Fe/H] and the absolute
magnitude of the stars $M_V$, using the HIPPARCOS parallax, for all elements
analysed (Fe, Ti, Ca, Ni, Cr, Si) for both neutral and ionised lines.

A particularly interesting example is Titanium. Looking at the case of TiI and TiII, we see
correlations, anti-correlations and negligible correlation for all three stellar
physical parameters probed. It turns out that very good fits for the measured
median EW difference can be made by simply fitting a plane as a function of
$T_{\mathrm{eff}}$ and [Fe/H], and only marginally improved fits made if we also include
$M_V$ as a linear term in the fitting:

\begin{equation} \label{eq:long}
\begin{split}
  \textless\Delta \mathrm{EW}_{\mathrm{n/i}}\textgreater = &
  \left. a(T_{\mathrm{eff}}-T_{\mathrm{eff},
    \astrosun}) + b([\mathrm{Fe/H}] - [\mathrm{Fe/H}]_{\astrosun})
  \right. \\ & \left. + c(M_{V} -
  M_{V, \astrosun}) \right. ,
\end{split}
\end{equation}

with $M_{V, \astrosun} = 4.83$ \citep{b22} and temperature
and metallicity values being those in the GCS-III catalogue.

This yields two equations for each element, one for the neutral lines and one
for the ionised lines. Setting the LHS of each equation to zero, we can solve
for any offset in the GCS temperature and metallicity calibrations.

The range of the absolute magnitude amongst our target stars is quite small ---
by design it is centred close to the Solar value --- and thus the range of
surface gravities in the stars (i.e log$(g)$) is expected to be very small.
Therefore we assume (and verify a posteriori, see below) its influence on our
spectral lines to be small. So, by neglecting the $M_{V}$
term in the above equations and rearranging, we find the two independent
relationships for effective temperature and metallicity:

\begin{equation} \label{eq:teff}
  T_{\mathrm{eff}} = T_{\mathrm{eff}, \astrosun} +
  a\textless\Delta \mathrm{EW}_{\mathrm{neutral}}\textgreater +
  b\textless\Delta \mathrm{EW}_{\mathrm{ionised}}\textgreater
\end{equation}

\begin{equation} \label{eq:metals}
  [\mathrm{Fe/H}] = [\mathrm{Fe/H}]_{\astrosun} + c\textless\Delta
  \mathrm{EW}_{\mathrm{neutral}}\textgreater + d\textless\Delta
  \mathrm{EW}_{\mathrm{ionised}}\textgreater
\end{equation}
which explicitly shows that the Solar values correspond to where both median
differences, in neutral and ionised lines, vanish.

For each of the 6 elements considered, a 2-D least squares fit to the data in
Fig.~2 and~3 provides us with the coefficients in Eq.~4 and~5; the intercepts
in particular correspond to the Solar values for temperature and metallicity
within the GCS.

Table \ref{tab:solval} shows the results we get for this 2-D fitting of the data.

\begin{table*}
 \centering
     \caption{Estimated Solar values for effective temperature and metallicity
       in the GCS calibration and the resulting $(b-y)$ colour of the Sun,
       determined through our (n/i) method.}
     \label{tab:solval}
  \scalebox{0.8}{
  \begin{tabular}{@{}lccccccccc@{}}
  \hline
   element & $T_{\mathrm{eff}, \astrosun}$ (K) & $a_{T_{\mathrm{eff}}}$ & $b_{T_{\mathrm{eff}}}$ & $[\mathrm{Fe/H}]_{\astrosun}$ (dex) & $c_{[\mathrm{Fe/H}]}$ & $d_{[\mathrm{Fe/H}]}$ & $(b-y)_{\astrosun}$ & $e_{(b-y)}$ & $f_{(b-y)}$\\
 \hline
 Iron & $5713\pm2$ & $-964\pm32$ & $1049\pm42$ & $-0.104\pm0.003$ & $1.29\pm0.05$ & $0.54\pm0.06$ & $0.4093\pm0.0002$ & $0.199\pm0.004$ & $-0.147\pm0.005$\\
 Titanium & $5717\pm2$ & $-906\pm31$ & $794\pm62$ & $-0.113\pm0.005$ & $1.12\pm0.07$ & $0.82\pm0.13$ & $0.4084\pm0.0002$ & $0.184\pm0.003$ & $-0.096\pm0.007$\\
 Calcium & $5748\pm5$ & $-1861\pm233$ & $444\pm125$ & $-0.093\pm0.006$ & $0.49\pm0.25$ & $0.86\pm0.14$ & $0.4042\pm0.0005$ & $0.338\pm0.024$ & $-0.064\pm0.015$\\
 Calcium 3D-fit&$5729\pm5$&&& $-0.110\pm0.006$ && & $0.4065\pm0.0007$ & &\\
 Nickel & $5707\pm4$ & $-658\pm64$ & $486\pm108$ & $-0.088\pm0.003$ & $0.83\pm0.06$ & $0.33\pm0.09$ & $0.4108\pm0.0005$ & $0.135\pm0.009$ & $-0.066\pm0.015$\\
 Chromium & $5728\pm2$ & $-1097\pm40$ & $1059\pm50$ & $-0.120\pm0.004$ & $1.25\pm0.06$ & $0.54\pm0.08$ & $0.4063\pm0.0003$ & $0.218\pm0.005$ & $-0.149\pm0.007$\\
 Silicon & $5706\pm3$ & $1233\pm100$ & $2492\pm157$ & $-0.105\pm0.003$ & $0.92\pm0.10$ & $0.11\pm0.15$ & $0.4103\pm0.0004$ & $-0.159\pm0.014$ & $0.394\pm0.023$\\
 \hline
 average & $5717\pm18$ & & & $-0.11\pm0.03$ & & & $0.409\pm0.002$ & &\\
\hline
\end{tabular} }
\end{table*}

We have tested the effect of neglecting the dependence of the EWs on absolute
magnitude in Eq.~3. This corresponds to including a linear term in
($M_{V} - M_{V, \astrosun}$) to
Eq.~\ref{eq:teff} and \ref{eq:metals}, and re-performing the least squares
fitting. We found the results were unchanged within the errors, along with
negligible improvement in the $\chi^{2}$ values of the fits. This shows that the dependencies on
$M_{V}$ in Fig.~\ref{fig:mv} is primarily driven by $T_{\mathrm{eff}}$
and [Fe/H] --- that is, once we account for these two dependencies, there is no
additional dependence of the $< \Delta EW >$ on magnitude/gravity, at least not
within the narrow selected magnitude range. There was an exception to this
result however: for Calcium, inclusion of an $M_{V}$ dependency improved the
fitting and provided zero--point values for temperature and metallicity more
consistent with the values from the other elements (see Table \ref{tab:solval}).

\begin{figure}
  \includegraphics[width=90mm]{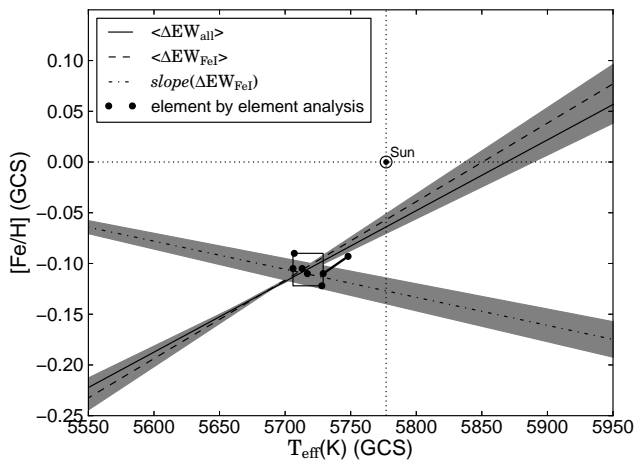}
  \caption[Optional caption for list of figures]{Filled circles show the
    resulting Solar temperatures and metallicites from the
    neutral-ionised-method.  The two connected points show the two results for
    Calcium. The lines show the trends from the degeneracy lines method, see
    Sec.~\ref{sub:deg}.  We also plot where the Sun lies in the GCS-III
    scale, as opposed to the location, where the degeneracy lines cross, which
    is where our method finds the most Sun-like stars in the sample.}
  \label{fig:finel}
\end{figure}

In Fig.~\ref{fig:finel} the filled circles show the GCS Solar temperature and
metallicity values we obtain for each of the 6 elements analysed.

We thus confirm, with a larger and higher quality dataset and a new method,
our results from Paper I: the GCS temperature scale is offset by
$-60\pm10$~K and the metallicity scale by $-0.11\pm0.02$~dex for Solar type
stars, which agrees with our previous results from Paper~I
within the errors ( $-100\pm40$~K and $-0.12\pm0.02$~dex respectively).

\subsubsection{The n/i-method: internal precision and performance using synthetic spectra}
\label{subsub:nitest}

As in Paper~I, our methods are novel and we have constructed several tests to
check their performance.

Our first test is to randomly chose 20 stars from the sample as reference
stars, instead of using Ceres (i.e. the Solar spectrum), to determine how well
the method recovers their catalogued temperatures and metallicities. We used
the same procedures as in section \ref{subsub:method}, but simply replace Ceres
with the randomly chosen star. Some examples of results for a few stars tested
are shown in Table \ref{tab:test}.

\begin{table}
 \centering
     \caption{Results from stars used to test the (n/i) method, showing their
       GCS temperatures and metallicities and how close we are to recovering
        those by applying our method to all the
       remaining stars in the sample. Note that this is a self-consistency
       check of the method only.}
     \label{tab:test}
  \begin{tabular}{@{}lcrrr@{}}
  \hline
   Name & $T_{eff}$ & $\Delta T_{\mathrm{eff}}$ & [Fe/H] & $\Delta$[Fe/H] \\
    & (GCS) & (fit-GCS) & (GCS) & (fit-GCS)\\
 \hline
 HD\,361    & 5821 & $-$7 & $-$0.25 & $-$0.02\\
 HD\,4391   & 5741 & 20 & $-$0.25 & 0.05\\
 HD\,13724  & 5675 & 5 & 0.01 & 0.04\\
 HD\,34449 & 5821 & $-$45 & $-$0.22 & 0.00\\
 HD\,59711 & 5715 & 6 & $-$0.21 & 0.02\\
 HD\,67458  & 5875 & $-$45 & $-$0.25 & $-$0.04\\
 HD\,78660 & 5715 & $-$27 & $-$0.09 & $-$0.01\\
 HD\,114853 & 5754 & $-$19 & $-$0.21 & $-$0.05\\
 HD\,126525 & 5585 & 84 & $-$0.19 & 0.07\\
 HD\,146233 & 5768 & $-$42 & $-$0.02 & $-$0.06\\
\hline
\end{tabular}
\end{table}

We recover the input temperatures and metalliticities of the stars very well,
finding an average temperature offset of only $\Delta T_{\mathrm{eff}}= -7 \pm40$~K and metallicity
offset of only $\Delta$[Fe/H]$=0.01\pm0.04$~dex between the input and output
values for the 10 stars. This demonstrates that the method is internally highly
consistent.

We have tried the same test using synthetic spectra rather than our
observational material. We used the POLLUX model spectra by \citet{b50},
selecting models which bracket the Solar values of temperature (5500~K, 5750~K
and 5600~K), metallicity ($-0.50$, $-0.25$, $0.00$, $+0.25$ and $+0.50$) and gravity (log(g)
= 4.0, 4.5 and 5.0). We arbitrarily adopt one of the spectra as our ``Solar''
reference spectrum, and determine if we can recover its temperature and metallicity
by applying our method to the rest of the stars.  The reference spectrum we
adopted was the model with $T_{\mathrm{eff}} = 5750$~K, [Fe/H] $= 0.00$~dex and log$(g) =
4.5$. Although these spectra span the Solar values rather widely and coarsely,
we nevertheless found our method recovers the correct temperature and
metallicity within the errors for the selected reference spectrum, yielding $T_{\mathrm{eff}} =
5757\pm18$~K and metallicity at [Fe/H] = $-0.03\pm0.03$~dex.

We conclude from these internal consistency tests that our (n/i) method
does indeed recover the parameters of the reference target.

\subsection{The degeneracy lines method applied to HARPS data}
\label{sub:deg}

For consistency and comparison to our previous work, we also apply our
``degeneracy lines'' methods (i) and (ii) from Paper~I.  Method (i) is based on
differential comparison of the EW's of 323 mostly neutral lines of 10 elements
(Ti, Fe, Al, Ca, Cr, Mg, Na, Ni, Si and Zn), and method (ii) combines
$\textless\Delta \mathrm{EW}_{\mathrm{Fe\thinspace I}} \textgreater$ and slope of
$\textless\Delta \mathrm{EW}_{\mathrm{Fe\thinspace I}} \textgreater$ versus 
excitation
potential of 129 Fe\thinspace I lines.  Because of the different telescopes and
instruments, the line list we used this time is different from the one we used
in Paper I. We decided not to use here the line depth versus excitation potential method 
(method (iii) in Paper I), which closely
resembled method (ii) but using line depths instead of
equivalent widths: we found in Paper I that these methods are virtually
indistinguishable, so we now consider method (iii) redundant.

\begin{figure*}
\centering
\subfigure{
  \includegraphics[scale=0.35]{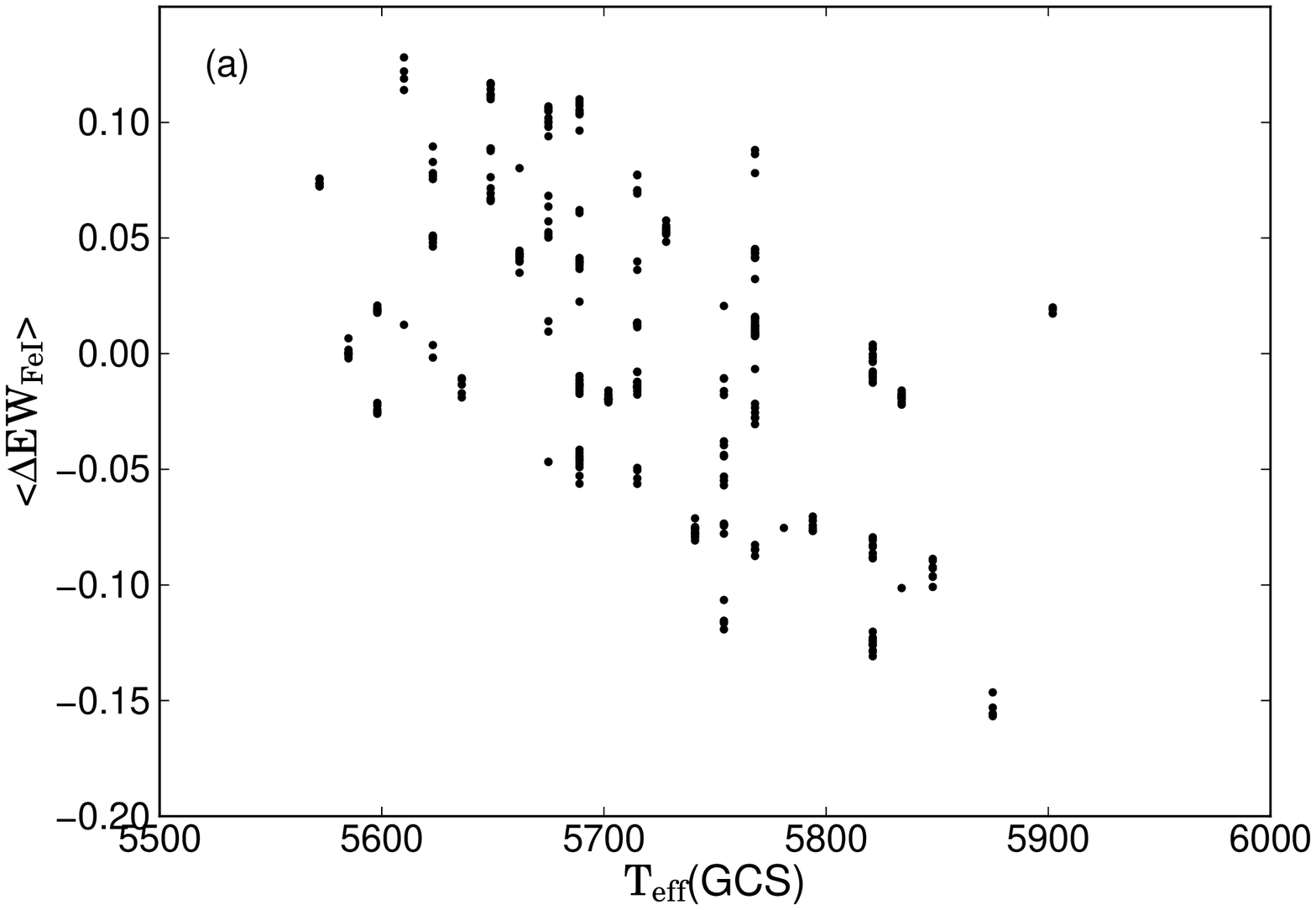}
\label{fig:subfig5}
}
\subfigure{
  \includegraphics[scale=0.35]{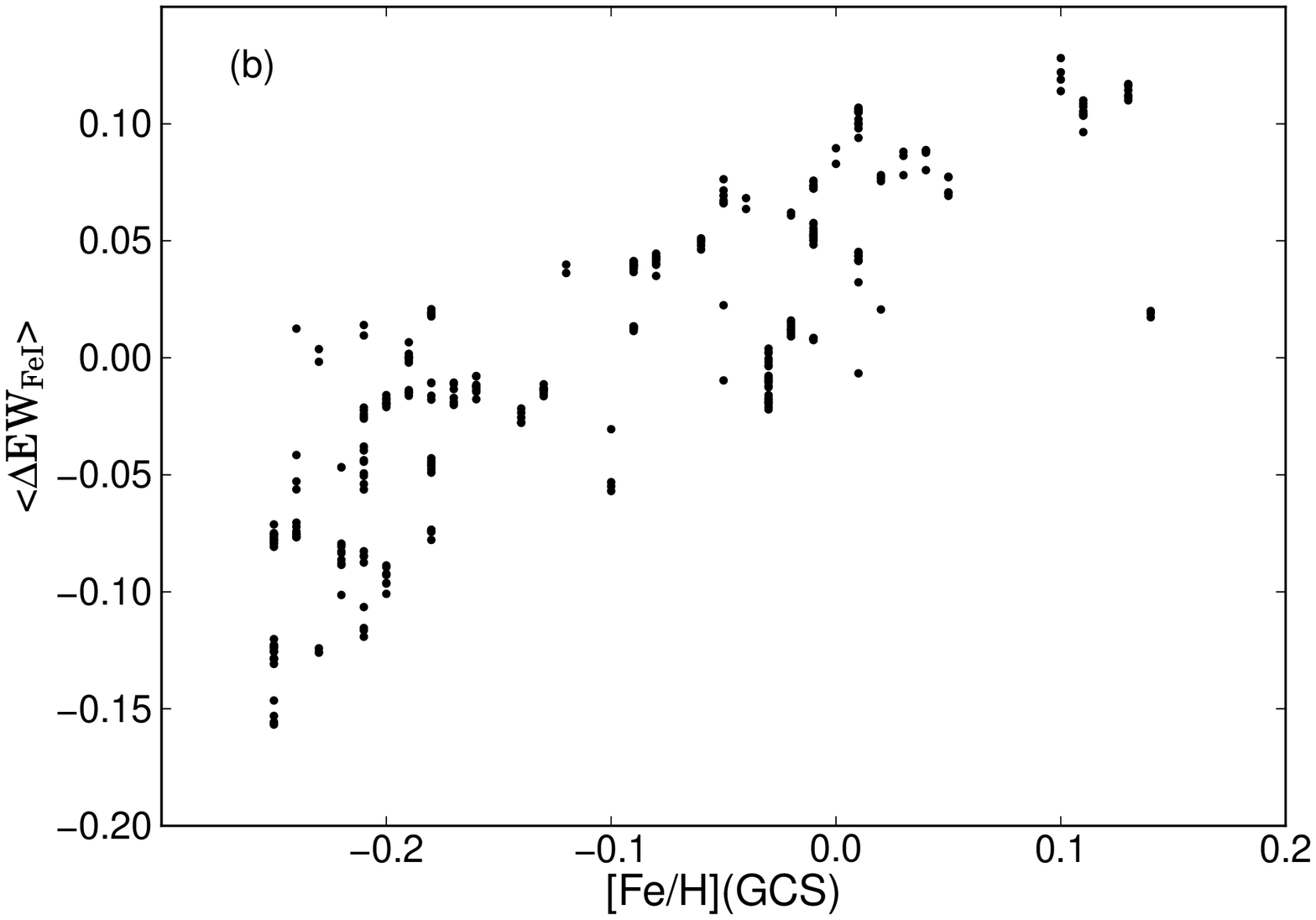}
\label{fig:subfig6}
}
\subfigure{
  \includegraphics[scale=0.35]{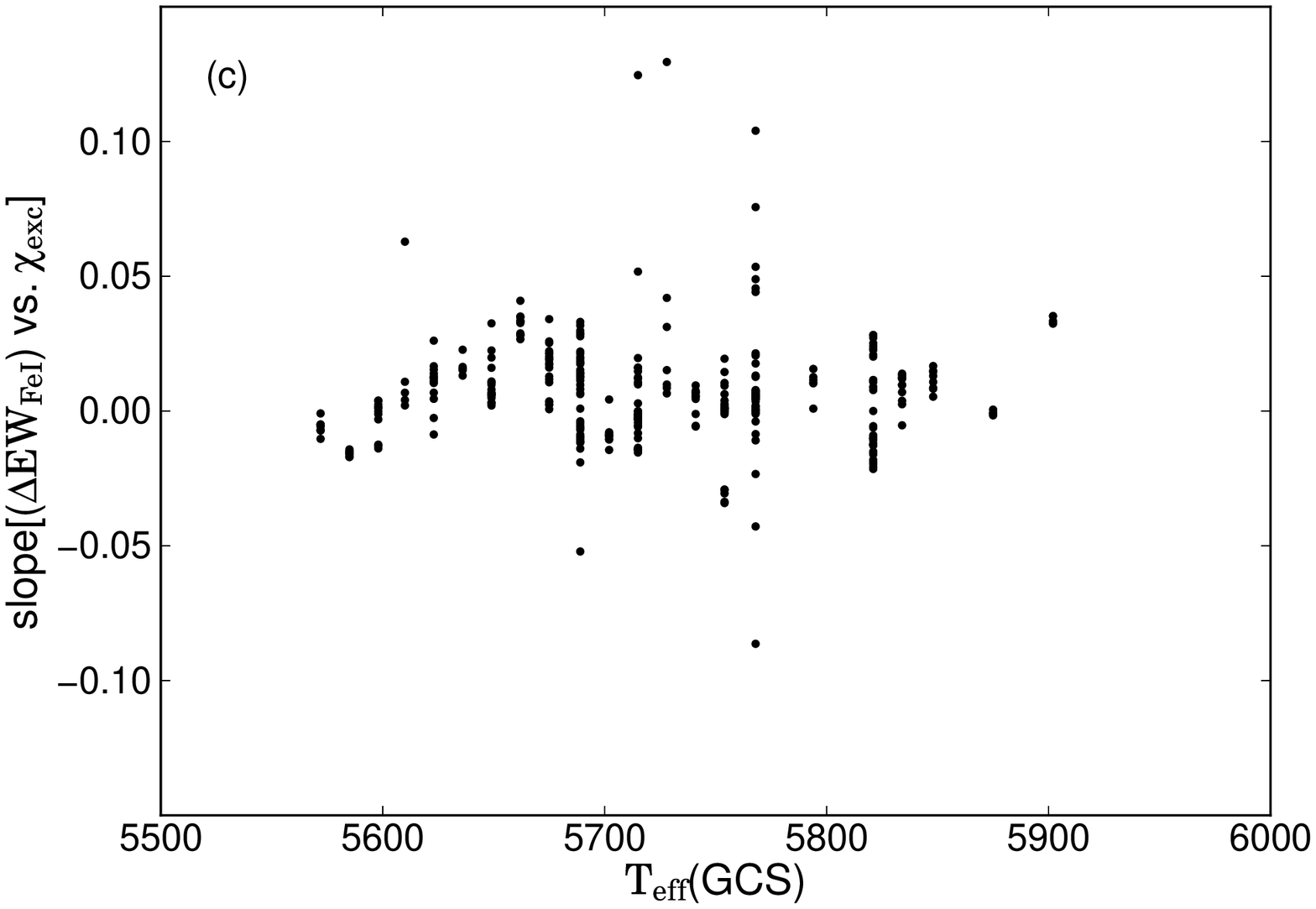}
\label{fig:subfig7}
}
\subfigure{
  \includegraphics[scale=0.35]{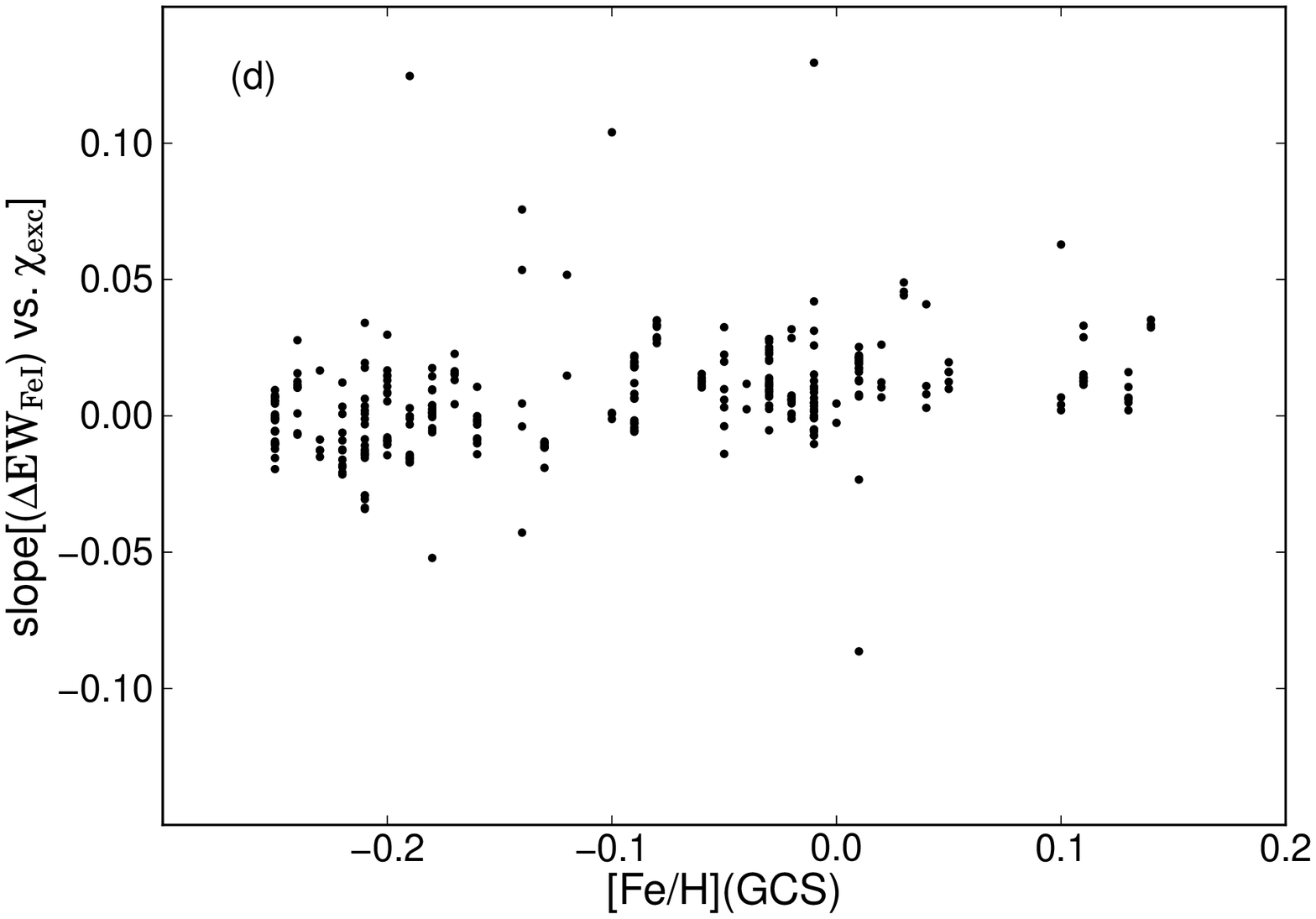}
\label{fig:subfig8}
}
\caption[Optional caption for list of figures]{Panel (a): the median difference
  in EW of the neutral iron lines for all target stars, depending on their
  temperatures and their metallicities (Panel (b)). Panel (c): the slopes of
  the relation between the median relative difference in EW as a function of
  the excitation potential of the line (details in Paper I); depending on
  temperature and (Panel (d)) metallicity. We use these trends in
  Section~\ref{sub:deg} to determine the offsets in temperature and
  metallicity in the GCS.}
\label{fig:dependancies}
\end{figure*}

We used our TWOSPEC code to determine the relevant quantities $\textless\Delta
\mathrm{EW}_{\mathrm{all}}\textgreater$, $\textless\Delta
\mathrm{EW}_{\mathrm{Fe\thinspace I}}\textgreater$ and slope[($\Delta
  \mathrm{EW}_{\mathrm{Fe\thinspace I}}$) vs. $\chi_{\mathrm{exc}}$].  We made a 2-D
least-squares fit of these quantities, to solve their combined dependancy on
metallicity and temperature (GCS values) --- see Fig.~\ref{fig:dependancies}:

\begin{equation} \label{eq:EWall}
\begin{split}
  \textless\Delta \mathrm{EW}_{\mathrm{all}}\textgreater = &
  ~0.493[\mathrm{Fe/H}]-1.986\left(\frac{T_{\mathrm{eff}}-5777}{5777}\right)
  \\ & +0.032 \\
\end{split}
\end{equation}

\begin{equation} \label{eq:EWfe}
\begin{split}
  \textless\Delta \mathrm{EW}_{\mathrm{Fe\thinspace I}}\textgreater = &
  ~0.511[\mathrm{Fe/H}]-2.284\left(\frac{T_{\mathrm{eff}}-5777}{5777}\right)
  \\ & +0.029 \\
\end{split}
\end{equation}

\begin{equation} \label{eq:slopeEWfe}
\begin{split}
  \mathrm{slope}[(\Delta
    \mathrm{EW}_{\mathrm{Fe\thinspace I}})\ \mathrm{vs.}\ \chi_{\mathrm{exc}}]= &
  ~0.519[\mathrm{Fe/H}] \\ &
  +0.829\left(\frac{T_{\mathrm{eff}}-5777}{5777}\right)+0.066 \\
\end{split}
\end{equation}

The left-hand side of the three equations vanish for a perfect Solar twin, so we can set
the left-hand side to zero and invert the equations, yielding:

\begin{equation} \label{eq:fehall}
  [\mathrm{Fe/H}]_{\textless\Delta \mathrm{EW}_{\mathrm{all}}\textgreater} =
  4.028\left(\frac{T_{\mathrm{eff}}-5777}{5777}\right)-0.064
\end{equation}

\begin{equation} \label{eq:fehfe}
  [\mathrm{Fe/H}]_{\textless\Delta \mathrm{EW}_{\mathrm{Fe\thinspace I}}\textgreater} =
  4.467\left(\frac{T_{\mathrm{eff}}-5777}{5777}\right)-0.057
\end{equation}

\begin{equation} \label{eq:slopefehfe}
\begin{split}
  [\mathrm{Fe/H}]_{\mathrm{slope}[(\Delta
      \mathrm{EW}_{\mathrm{Fe\thinspace I}})\ \mathrm{vs.}\ \chi_{\mathrm{exc}}]} = &
  -1.598\left(\frac{T_{\mathrm{eff}}-5777}{5777}\right) \\ & -0.127 \\
\end{split}
\end{equation}

As in Paper I, we found the slope parameter to be almost completely independent
of temperature (see Fig.~\ref{fig:subfig5}) although we formally kept the
dependence in the equation, for consistency with the analysis of the scale of \citet{b40}
in section~\ref{sub:revGCS}.

Plotting these relations in the metallicity-temperature plane, we find, as in
Paper I, that they cross at a point offset from the Solar values, by $-0.10
\pm 0.02$~dex in
metallicity and $-50 \pm 25$~K in temperature (see Fig.~\ref{fig:finel}).

It is possible that the strength of the our spectral lines may be 
introducing a bias into
our method: to check this, we divided the lines into median strength 
($log(\frac{EW}{\lambda})>-5$), and weak ($log(\frac{EW}{\lambda})<-5$);
each sublist contains about half of the lines of the original full list.
We find no change to the results when adopting each sublist in turn.
Our list of solar twins, selected in Section 4 with similar methods,
is also robust to this test.

\subsection{The reanalysis of the GCS (C11)}
\label{sub:revGCS}

\citet{b40} have analysed the GCS catalog, using the InfraRed Flux Method on
Solar twins to set the temperature scale, and redetermining the physical
parameters of the stars. They find their calibration to be $\sim$80~K hotter and
$\sim$0.1~dex metal richer than in the GCS.

\begin{figure}
  \includegraphics[width=90mm]{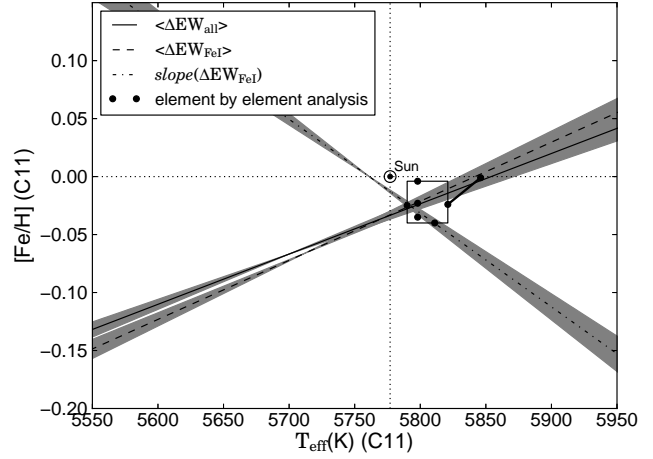}
  \caption[Optional caption for list of figures]{Same as Fig.~\ref{fig:finel},
    but using C11 values for temperature and metallicity. Again we
    also plot the position of the Sun in the C11 scale.}
  \label{fig:luca}
\end{figure}

We have applied our method to their temperatures and metallicities, to check
the zero point of their scales relative to the Sun. We apply all the methods
discussed above, and show the results in Fig.~\ref{fig:luca}.

We find the Solar values in C11 are at
$T_{\mathrm{eff}}=5790\pm15$~K and
$[\mathrm{Fe/H}]=-0.02\pm0.02$~dex: this is very close to the accepted Solar
values, and so this work favours the C11 scales over the original one,
at least for Sun--like stars (i.e. within a window of 250~K in temperature and
0.15~dex in metallicity around the Sun). We note this is not particularly
surprising considering that the temperature scale of \citet{b24,b40} was
explicitly calibrated on Solar twins.

\subsection{The Solar $(b-y)$ colour}

Analogous to Paper~I, we applied the same procedures using the $(b-y)$ colour of
the target stars instead of the temperature, to determine the Solar $(b-y)$
colour.  For this purpose we first fitted the following relation to our six
different elements, as in our (n/i)-method, see Sec.~\ref{sub:nim}:

\begin{equation} \label{eq:by}
  (b-y) = (b-y)_{\astrosun} + e\textless\Delta
  \mathrm{EW}_{\mathrm{neutral}}\textgreater + f\textless\Delta
  \mathrm{EW}_{\mathrm{ionised}}\textgreater
\end{equation}

\begin{figure}
  \includegraphics[width=90mm]{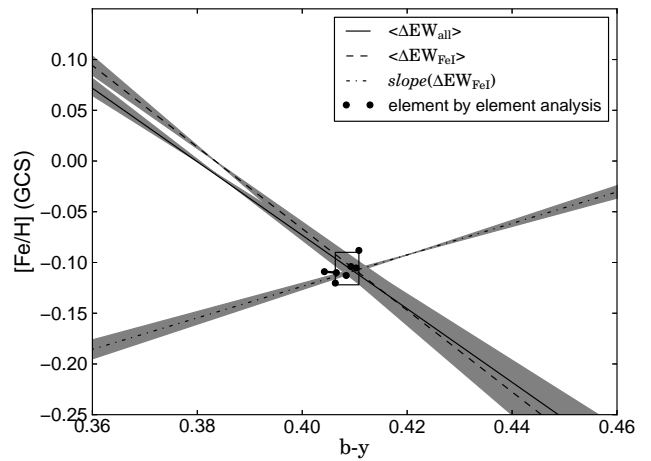}
  \caption[Optional caption for list of figures]{Same as Fig.~\ref{fig:finel},
    but looking at the $(b-y)$ colour instead of the temperature. The Solar
    colour is derived from the crossing point of the relations and yields a
    $(b-y)_{\astrosun}$ colour of $0.409\pm0.005$.}
  \label{fig:finalby}
\end{figure}

We then used the degeneracy lines method, as in Sec.~\ref{sub:deg}, and derived
an overall estimate for the Solar $(b-y)$ colour of $0.409\pm0.005$
(Fig.~\ref{fig:finalby} and Table~\ref{tab:solval}).

This is very consistent with the results of \citet{b23}, who obtained
$(b-y)=0.4105\pm0.0015$ through Str{\"o}mgren photometry and also with the
value we found in Paper~I: $(b-y)=0.414\pm0.007$.

\section{Finding the Solar twins from HARPS}
\label{sec:twins}

As shown in Paper~I, there are currently many different ways of selecting Solar
twins. The interested reader is referred to it for full details; we give short
descriptions of each method in what follows.

We firstly use the median difference in equivalent width of a list of spectral
lines (Paper I, method (i)) and the median difference in equivalent width of
only the Fe\thinspace I lines (Paper I, method (ii)).

We have also adapted the (n/i)-method from section \ref{sub:nim} of this paper, as
another way to look for Solar twins. This relies on finding the minimum of the
median difference in equivalent width of specific elements (Fe, Ti, Ca, Ni, Cr,
Si) but separating the lines from the neutral species and the ones from the
singularly ionised species; see method (n/i) (Section~\ref{sub:meth3}).

\subsection{Method (i): the equivalent widths of all 321 spectral lines}
\label{sub:meth1}

As shown in Paper~I, this method (which is similar to the ``first criterion''
method of \citet*{b3}), uses the median
$\textless\Delta\mathrm{EW}_{\mathrm{all}}\textgreater$ and scatter
$\chi^{2}(\Delta \mathrm{EW}_{\mathrm{all}})$ of the differences
$\Delta$EW$_{\mathrm{all}}$ in the EWs of target stars relative to Ceres for
all our lines, to determine a star's Solar likeness.

To determine the errors in the measurements of our various parameters in
methods (i) and (ii), we used the fact that we had many targets with several
spectra. Comparing the measurements with one another, resulted in the following
error values: $\sigma(\chi^{2}(\Delta \mathrm{EW}_{\mathrm{all}}))$=0.1,
$\sigma(\textless\Delta \mathrm{EW}_{\mathrm{all}}\textgreater\nolinebreak)$=0.001,
$\sigma(\textless\Delta \mathrm{EW}_{\mathrm{Fe\thinspace I}}\textgreater)$=0.001 and
$\sigma$(slope($\Delta \mathrm{EW}_{\mathrm{Fe\thinspace I}}$)
vs. $\chi_{\mathrm{exc}}$)=0.001.

\begin{figure}
  \includegraphics[width=90mm]{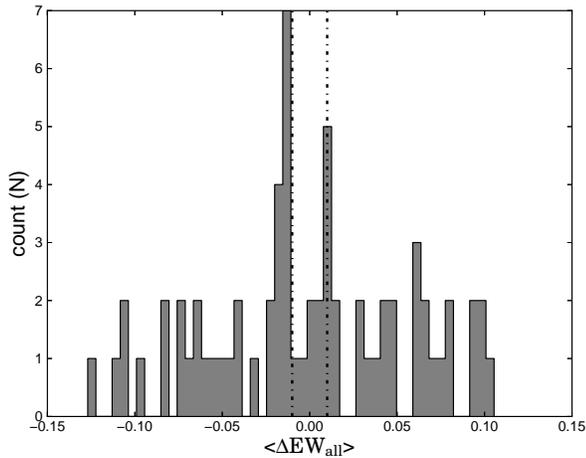}
  \caption[Optional caption for list of figures]{Histogram of the median
    relative difference in EWs for all lines
    $\textless\Delta\mathrm{EW}_{\mathrm{all}}\textgreater$ in our sample. The
    broken lines show where we applied the 1\% cuts, thus every stellar spectrum
    within the lines is considered to be that of a Solar twin.}
  \label{fig:hist}
\end{figure}

In Paper I the definition
of a solar twin has been based on a match to the Solar spectrum to within a
limit defined by the observational scatter. We used the criterion
that our twins should be Solar within 2-$\sigma$, where $\sigma$ was the
observational scatter, as determined from multiple measurements of the
indicator used in each method. Our HARPS data are now so accurate, that the
same 2-$\sigma$ criterion yields no Solar twins. We therefore adopt the
following expedient: we define Solar twins as being within 1\% of the Sun in
the measured indicator. For the median method, the difference between the EW of
the lines in the star and in the Sun should differ by less than 1 percent of
the Solar value. Fig.~\ref{fig:hist} shows an overview on the range of values
and the twin selection limits for this criterion.

Additionally we define the scatter to be only the observational scatter, in this case 
$\chi^{2} \leq 1$. Thus we find 5 stars we consider to be Solar
twins, as shown in Table~\ref{tab:chi2}. The well known twin 18\,Sco lies just outside the
limits with a $\textless\Delta\mathrm{EW}_{\mathrm{all}}\textgreater$ of 1.2\%
of Solar.

\begin{table}
 \centering
     \caption{List of Solar twins using $\chi^{2}(\Delta
       \mathrm{EW}_{\mathrm{all}})$ (i.e. method (i)), ordered by $\chi^{2}(\Delta
       \mathrm{EW}_{\mathrm{all}})$.}
     \label{tab:chi2}
  \begin{tabular}{@{}lcr@{}}
  \hline
   Name & $\chi^{2}(\Delta \mathrm{EW}_{\mathrm{all}})$ 
        & \myalign{c}{$\textless\Delta \mathrm{EW}_{\mathrm{all}}\textgreater$}\\
 \hline
 HD\,19641  & $0.5\pm0.1$ & $-0.005\pm0.001$ \\
 HD\,78660  & $0.5\pm0.1$ & $ 0.006\pm0.001$ \\
 HD\,45184  & $0.6\pm0.1$ & $ 0.001\pm0.001$ \\
 HD\,126525 & $0.9\pm0.1$ & $ 0.000\pm0.001$ \\
 HD\,76440  & $1.0\pm0.1$ & $ 0.006\pm0.001$ \\
\hline
\end{tabular}
\end{table}

\subsection{Method (ii): the equivalent widths of 129 Fe\thinspace I lines versus their 
excitation potential}
\label{sub:meth2}

This method, which we also adopted in Paper~I is originally inspired by the
technique used by \citet{b11}, where we use only the 129 Fe\thinspace I lines in our
line list. We determine the median $\textless\Delta
\mathrm{EW}_{\mathrm{Fe\thinspace I}}\textgreater$ for these lines and the
slope[($\Delta\mathrm{EW}_{\mathrm{Fe\thinspace I}}$) vs. $\chi_{\mathrm{exc}}$], which
should be zero for a Solar twin.

Analogously to the previous section, we adopt the definition of a star being
Solar, when these values lie within 1\% of the Solar values.

This gives us also five stars we can consider as Solar twins, see
Table~\ref{tab:meth2}, two of which also satisfied method (i): HD45184 and
HD76440.

\begin{table}
\centering
\caption{List of Solar twins from method (ii).}
\label{tab:meth2}
\begin{tabular}{@{}lrr@{}}
\hline
Name & \myalign{c}{$\textless\Delta \mathrm{EW}_{\mathrm{Fe\thinspace I}}\textgreater$} 
     & \myalign{c}{slope[($\Delta \mathrm{EW}_{\mathrm{Fe\thinspace I}}$) vs. $\chi_{\mathrm{exc}}$]}\\
\hline
HD\,45184  & $-0.009\pm0.001$ & $ 0.009\pm0.001$ \\
HD\,76440  & $-0.001\pm0.001$ & $-0.009\pm0.001$ \\
HD\,78538  & $-0.008\pm0.001$ & $-0.007\pm0.001$ \\
HD\,146233 & $ 0.009\pm0.001$ & $ 0.007\pm0.001$ \\
HD\,183658 & $ 0.008\pm0.001$ & $ 0.001\pm0.001$ \\
\hline
\end{tabular}
\end{table}

\subsection{Method (n/i)}
\label{sub:meth3}

In addition to the previous criteria, which are the same as in Paper~I, we
introduce the new criterion inspired by our neutral/ionised lines method of
Section~\ref{sub:nim}, by considering a star to be a Solar twin, if, for more
than one element used in our analysis (Fe,
Ti, Ca, Ni, Cr and Si), see Table~\ref{tab:meth3}, the following criterion holds:

\begin{equation} \label{eq:ni}
 |\textless\Delta \mathrm{EW}_{\mathrm{neutral}}\textgreater| +
 |\textless\Delta \mathrm{EW}_{\mathrm{ionised}}\textgreater| \leq 0.02
\end{equation}

\begin{table}
\centering
\caption{List of Solar twins from method (n/i).}
\label{tab:meth3}
\begin{tabular}{@{}lc@{}}
\hline
Name & elements for which the criterion is fulfilled\\
\hline
HD\,197027 & Fe, Ni, Cr, Si\\
HD\,76440  & Fe, Ca, Ni, Si\\
HD\,78660  & Ca, Ni, Si\\
HD\,19641  & Ca, Cr\\
HD\,126525 & Ca, Ni\\
HD\,146233 & Ti, Ca\\
\hline
\end{tabular}
\end{table}

This yields six twins, five of which have already been found to be twins in the
previous sections, which shows the analysis is robust to changes in the twin
definition. The sixth twin was only recently found to be the oldest twin known
to date by \citet{b58} at an age of $\sim 8.2$~Gyrs.

When comparing the abundances of different elements in the Sun to those 
in solar twins, \citet{b60} showed that volatile elements are more abundant in the Sun, 
whereas refractory elements are either of similar or lower abundance.
In our study this is in principle a concern for silicon, a volatile element whose
relative abundance in \citet{b60} is found to be about $0.03$~dex higher 
than that of iron or of other refractory elements like Ti.
In practice, we find no sign of this dichotomy in our results: in Section \ref{subsub:method}
there is no significant difference between the Si--based results of the n/i 
method versus other elements; and the solar twins we selected with the various 
methods are not peculiar, and very close to solar, in $< \Delta EW_{Si}>$.
We plan to return to this issue in the near future, with a detailed abundance 
analysis of our twins (from this paper and Paper I) based on UVES spectra.

\subsection{Final list of Solar twins}
\label{sub:final}

Looking at all three methods together, we classify the following stars as the
Solar twins in our sample (Table~\ref{tab:final}).

\begin{table*}
 \centering
     \caption{Our Solar twins compared to the Sun. Note that the quoted Solar
       $(b-y)$ is estimated indirectly from Sun-like stars by
       \citet{b19}. Stars marked with (*) may have poor temperatures, due to
       overestimated reddening corrections (see Section \ref{sub:final}).}
     \label{tab:final}
  \begin{tabular}{@{}lrrrrrrlc@{}}
  \hline
   Name & $(b-y)$ & $M_{V}$ & [Fe/H] & [Fe/H]& $T_{\mathrm{eff}}$ (K) & $T_{\mathrm{eff}}$ (K) & selection\\
   &&& (GCS) & (C11) & (GCS) & (C11) & method\\
 \hline
       Sun (Holmberg) & 0.403 & 4.83 & 0.00 && 5777 &&\\
       Sun (this work) & 0.409&&&&&&\\
\hline
\hline
    HD\,76440$^{*}$ & 0.419 & 4.82 & $-$0.23 & 0.02 & 5623 & 5991 & i, ii, n/i\\
\hline
    HD\,19641$^{*}$  & 0.407 & 4.67 & $-$0.19 & 0.00 & 5715 & 5928 & i, n/i\\
    HD\,45184  & 0.394 & 4.67 & $-$0.03 & 0.04 & 5821 & 5863 & i, ii\\
    HD\,78660  & 0.409 & 4.75 & $-$0.09 & $-$0.03 & 5715 & 5788 & i, n/i\\
    HD\,126525 & 0.426 & 4.94 & $-$0.19 & $-$0.10 & 5585 & 5666 & i, n/i\\
    HD\,146233  & 0.404 & 4.79 & $-$0.02 & 0.06 & 5768 & 5826 & ii, n/i\\
\hline
    HD\,78538 & 0.407 & 4.95 & $-$0.16 & $-$0.09 & 5715 & 5781 & ii\\
    HD\,183658 & 0.403 & 4.80 & $-$0.01 & 0.06 & 5768 & 5826 & ii\\
    HD\,197027 & 0.421 & 4.72 & $-$0.24 & $-$0.17& 5610 & 5774 & n/i\\
\hline
    average (all) & 0.410 & 4.79 & $-$0.13 & $-$0.02 & 5702 & 5827&\\
    average (w/o dereddened stars) &&& $-$0.11 & $-$0.03 & 5711 & 5789 &\\
\hline
\end{tabular}
\end{table*}

Only one star in our sample satisfies all three of our criteria: HD\,76440.
There are five stars which satisfy two of the criteria and three stars that
satisfy one criterion.

Of the ten Solar twins we reported in Paper~I from the FEROS data, three were
found in the HARPS archive and included in our sample for this paper. All three
here are also identified as twins, using the methods described above, which
makes us confident in the robustness of our methods' end results, as we
recovered the same, common twins with a different telescope, instrument, line
list and method.  These three stars are HD\,78660, HD\,126525 and HD\,146233.

Of the nine stars we consider Solar twins here, two have been previously
published as twins in and before Paper~I, them being HD\,146233 and HD\,78660;
HD197027, was only recently found to be an old twin \citep{b58}; the other six
stars are completely new twins: HD\,45184, HD\,76440, HD\,19641, HD\,78538 and
HD\,183658.

\begin{figure}
  \includegraphics[width=90mm]{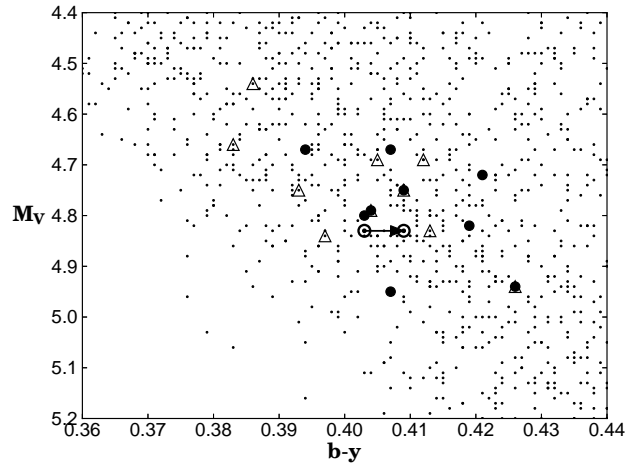}
  \caption[Optional caption for list of figures]{The colour-magnitude-diagram
    for the GCS entries around the Solar values \citep{b9}. The open triangles 
    show the position of the Solar twins we found in Paper~I, the filled circles 
    are the twins from this study. The Sun is marked by the Sun
    symbol, using the colour values by \citet{b19} for the left one, whereas
    the arrow is pointing to the position using our own determined value for
    $(b-y)$.}
  \label{fig:twins}
\end{figure}

In Fig.~\ref{fig:twins} we show where the twins lie in the
colour-magnitude-diagram, compared to the twins from Paper~I.

To check to what extent the choice of line list used influences our results, we
tested the effect of using the shorter FEROS line list from Paper~I on the
HARPS data, and also taking into account the reduced wavelength coverage of the
HARPS spectra. Using this line list, with $\sim 100$ lines, we
recover four of the nine twins mentioned above, namely HD\,45184, HD\,76440,
HD\,126525 and HD\,183658. Our methods are quite robust, as we recover the same
twins for a broad range of comparison lines, species, ionisation states,
telescopes and instrumental resolutions.

Using the average
properties of Solar twins to check calibrations is a widely applied technique
\citep{b53, b24}.  We do the same here as a consistency check, but consider
this not as robust as our previously described methods.  When using all twins
we obtain for the GCS-III an average metallicity of $-0.13\pm0.09$~dex and
temperature of $5702\pm80$~K, which confirms the previously found offsets
in Section \ref{sec:analysis}.  Notice that two of our twins in
Table~\ref{tab:final} show very hot temperatures in C11.  They have
been marked (*) to have reddening that is not insignificant ($E(B-V) > 0.01$~mag). The GCS
assigns them to be 0.044~mag and 0.027~mag, respectively, despite the fact that
they are only 50~pc away. Therefore it is possible that their reddening
might have been overestimated for the C11 catalogue. A difference in reddening of  
$\Delta E(B-V) = 0.01$ corresponds to a change in temperature of 
$\Delta T_{\mathrm{eff}}$ = 30-50~K. If the stars are negligibly reddened, as their 
proximity would suggest, they are likely to be up to 200~K cooler than in C11. 
If we take out these two stars from the average, the new values are $5711\pm86$~K 
for the GCS and $5789\pm63$~K for C11, thus still favouring the latter calibration.

\section{Summary and conclusions}
\label{sec:sum}

We have used several hundred weak neutral and ionised absorption lines for a
range of atomic species to search for stars as similar to the Sun as possible
(``Solar twins'') by comparing their spectra to a Solar reflection spectrum of
the asteroid Ceres.

In Paper~I, we found offsets in the temperature and metallicity scales of the
Geneva-Copenhagen Catalog (GCS), at least for Solar type stars, by using data
from the FEROS instrument on the MPG/ESO 2.2m telescope.  Here we confirm the
offsets, using more and considerably higher quality data, taken with the HARPS
instrument on the 3.6m ESO telescope. This makes us confident that these
offsets are real, which are $-55 \pm 25$~K in the temperature and $- 0.10\pm 0.03$~dex  
in the metallicity scale for Solar type
stars. These offsets are somewhat smaller, especially for temperature,
than those in Paper I; but they agree within the errors. Notice that one method we used
in Paper I (the line depth ratio method) also indicated a smaller offset in temperature
of about 50~K. 

Note that all the errors quoted in the conclusions
are averaged from the range of results among the different methods 
applied in this paper, which are discussed separately in the corresponding sections
and are comparable to the internal errors of each method.

Our analysis favours the temperature and metallicity scale of \citet[C11]{b40} 
--- explicitly tuned on Solar twins --- over
the GCS-III values. We find small offsets of $-13\pm15$~K in temperature and $-0.02\pm0.02$~dex in
metallicity around the Solar values, which are within the error limits of our
method.

In principle, the most fundamental test of the temperature scale is comparison 
with interferometry, but extant interferometric data do not yet allow for a 
strong conclusion about the GCS-III/C11 temperature calibration 
\citep{b61}: the agreement of the two scales with interferometry 
is within 50~K or better, on the cool and hot side respectively; the precise
value of the offsets depending on the specific sample of interferometric stars 
considered for the comparison.
Therefore, alternative tests of the temperature calibration remain useful. The one 
we presented in this paper and in Paper I has the advantage of relying on systematic 
trends established over a significant number of Sun-like stars.

From the HARPS data we also estimate the Solar $(b-y)$ colour to be
$0.409\pm0.002$, consistent with the $(b-y)=0.403\pm0.013$ found by
\citet{b19}, and close to what \citet{b23} found: $(b-y)=0.4105\pm0.0015$ and
also close to what we found in Paper~I: $(b-y)=0.414\pm0.007$.

We also used this new set of data to further focus on finding Solar twins. We
found five new ones (HD\,19641, HD\,45184, HD\,76440, HD\,78538 and
HD\,183658), one that was only recently the focus of attention: HD\,197027
\citep{b52} and confirmed all three twins in common with the sample of Paper~I
(HD\,78660, HD\,126525 and HD\,146233/18~Sco), guaranteeing consistency with
our previous results.

As in Paper~I, one of our best twins is HD\,126525. This is a puzzle to us, as
its temperature and metallicity values from the literature are very different
from Solar and our internal precision testing assigns
to it values that are off by 100~K and 0.12~dex. We are currently analysing
very high resolution and high S/N VLT/UVES spectra of this target to have a
closer look at its elemental abundances to better settle the nature and
parameters of this intriguing twin. It is also one of our targets which has a
confirmed planetary companion \citep{b31}.

We demonstrate that the use of Solar twins and Solar analogues is an accurate
and precise means of testing the stellar temperature and metallicity scales for
sun-like stars. Our methods in this paper and Paper~I show a high degree of
consistency and reliability, but have been applied so far only on the GCS
catalogue. We plan to use these to test the calibrations of other scales for
which temperature and metallicity estimates for Solar analogues are
available. It should be straight-forward to extend the applicability of our
method beyond Sun-like stars, providing a way to probe any metallicity and
temperature regime, by reference to a set of stars with accurate determinations
of their fundamental parameters. In the past few years interferometric surveys
have started to become that source of stars with parameters in the fundamental
scale.  Using these to anchor our methods to, we plan to
probe other regions in the temperature-metallicity plane to provide a more
global check of the calibration in spectro--photometric catalogues.

\section*{Acknowledgments}

We are grateful to Johan Holmberg for originally suggesting to us
the neutral/ionized method presented in this paper. We also thank
our referee, Bengt Gustafsson, for constructive remarks.
This study was financed by the Academy of Finland (grant nr.~130951 and 218317)
and the StarryStory Fund. We thank Swinburne
University, where part of this work was carried out.

\bsp

\label{lastpage}

\end{document}